\documentclass[%
reprint,
superscriptaddress,
 amsmath,amssymb,
 aps,
prl,
longbibliography,
]{revtex4-2}

\usepackage{graphicx}
\usepackage{dcolumn}
\usepackage{bm}
\usepackage{braket}
\usepackage{amsmath}
\usepackage{xcolor}
\usepackage[normalem]{ulem}
\usepackage[colorlinks, citecolor={blue!80!black}, urlcolor={blue!80!black}, linkcolor={red!60!black}]{hyperref}
\definecolor{myblue}{RGB}{0, 112, 192}

\begin{document}

 \author{L.~Banszerus}
 \thanks{Present address: University of Vienna, Faculty of Physics, Boltzmanngasse 5, 1090 Vienna, Austria}
 \thanks{These authors contributed equally to this work}
 \affiliation{JARA-FIT and 2nd Institute of Physics, RWTH Aachen University, 52074 Aachen, Germany,~EU}%
 \affiliation{Peter Gr\"unberg Institute  (PGI-9), Forschungszentrum J\"ulich, 52425 J\"ulich,~Germany,~EU}

 \author{K.~Hecker}
 \thanks{These authors contributed equally to this work}
 \affiliation{JARA-FIT and 2nd Institute of Physics, RWTH Aachen University, 52074 Aachen, Germany,~EU}%
 \affiliation{Peter Gr\"unberg Institute  (PGI-9), Forschungszentrum J\"ulich, 52425 J\"ulich,~Germany,~EU}

 \author{L.~Wang}
 \affiliation{Department of Physics, University of Konstanz, 78457 Konstanz, Germany,~EU}%

 \author{S.~M\"oller}
 \affiliation{JARA-FIT and 2nd Institute of Physics, RWTH Aachen University, 52074 Aachen, Germany,~EU}%
 \affiliation{Peter Gr\"unberg Institute  (PGI-9), Forschungszentrum J\"ulich, 52425 J\"ulich,~Germany,~EU}

 \author{K.~Watanabe}
 \affiliation{Research Center for Electronic and Optical Materials, National Institute for Materials Science, 1-1 Namiki, Tsukuba 305-0044, Japan}
 
 \author{T.~Taniguchi}
 \affiliation{Research Center for Materials Nanoarchitectonics, National Institute for Materials Science,  1-1 Namiki, Tsukuba 305-0044, Japan}

 \author{G.~Burkard}
 \affiliation{Department of Physics, University of Konstanz, 78457 Konstanz, Germany,~EU}%

 \author{C.~Volk}
 \author{C.~Stampfer}
 \affiliation{JARA-FIT and 2nd Institute of Physics, RWTH Aachen University, 52074 Aachen, Germany,~EU}%
 \affiliation{Peter Gr\"unberg Institute  (PGI-9), Forschungszentrum J\"ulich, 52425 J\"ulich,~Germany,~EU}%

\title{Phonon-limited valley lifetimes in single-particle bilayer graphene quantum dots}  

\date{\today}

\begin{abstract}
The valley degree of freedom in 2D semiconductors, such as gapped bilayer graphene (BLG) and transition metal dichalcogenides, is a promising carrier of quantum information in the emerging field of valleytronics.
While valley dynamics have been extensively studied for moderate band gap 2D~semiconductors using optical spectroscopy techniques, very little is known about valley lifetimes in narrow band gap BLG, which is difficult to study using optical techniques. 
Here, we report single-particle valley relaxation times ($T_1$) exceeding several microseconds in electrostatically defined BLG quantum dots using a pulse-gating technique. 
The observed dependence of $T_1$ on perpendicular magnetic field can be understood qualitatively and quantitatively by a model in which $T_1$ is limited by electron-phonon coupling. 
We identify the coupling to acoustic phonons via the bond length change and via the deformation potential as the limiting mechanisms.
\end{abstract}
\maketitle

Charge carriers in 2D materials with a hexagonal crystal lattice have, in addition to the spin, a tunable valley degree of freedom.
This renders these materials promising candidates for valleytronics~\cite{Rycerz2007Mar,Schaibley2016Aug,Bussolotti2018Jun,Mrudul2021Mar}, where the valley relaxation time ($T_1$) is an important figure of merit, allowing to assess the potential for valley-based information storage. 
In 2D transition metal dichalcogenides with moderate band gaps, the valley degree of freedom is well accessible for optical manipulation and readout. This allows pump-probe spectroscopy experiments~\cite{Kim2017Jul,Mai2014Jan}, in particular time-resolved Kerr rotation~\cite{Hsu2015Nov,Dey2017Sep,Yang2015Oct,Yan2017Jun,Zhao2021Mar}, which revealed relaxation times ranging from nanoseconds to microseconds depending on the material and excitation conditions. 
However, such optical techniques are not readily applicable to narrow-gap BLG, resulting in a lack of knowledge about valley lifetimes in BLG. 
Recent advances in the confinement of carriers in BLG using quantum dots (QDs), however, are opening up new avenues for the study of valley lifetimes in BLG.

\begin{figure}[ht]
\centering
\includegraphics[draft=false,keepaspectratio=true,clip,width=\linewidth]{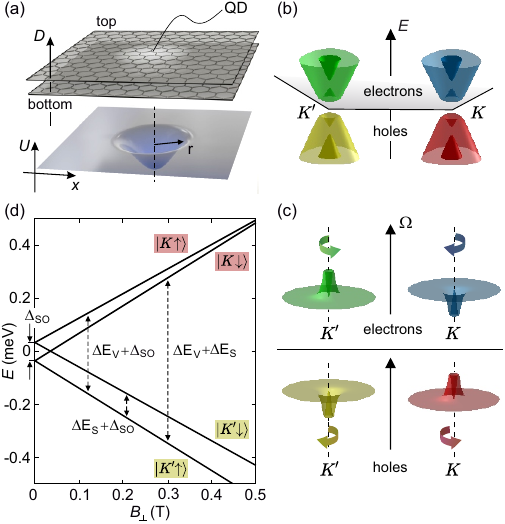}
\caption[Fig01]{
(a) Lattice structure of BLG highlighting the top and bottom layer and a symmetry breaking displacement field, $D$. The electrostatic confinement potential $U$ allows to form a QD.
(b) Band structure of gapped BLG. Close to the $K$ and the $K'$ valley a band gap opens.
(c) The broken inversion symmetry results in a finite Berry curvature, $\Omega$, near the $K$ and $K'$ points, which has opposite sign for the two valleys and for electrons and holes. 
(d) Single-particle spectrum of a BLG QD. At $B_\perp = 0$,
$\Delta_\text{SO}$ leads to the formation of two Kramers' doublets. A finite $B_\perp$ results in a spin and valley Zeeman effect, leading to additional energy splittings of $\Delta E_\text{s} = g_\mathrm{s}\mu_\mathrm{B}B$ and $\Delta E_\text{v} = g_\mathrm{v}\mu_\mathrm{B}B$. 
The arrows depict the transition energies between the ground state $\ket{K'\uparrow}$ and the three excited states.}
\label{f1}
\end{figure}

Bernal-stacked BLG comes as a gapless semimetal, where electrons and holes can be described as massive Dirac fermions~\cite{Novoselov2006Mar}. However, when applying an out-of-plane electric displacement field $D$, the inversion symmetry of the crystal lattice is broken, as the on-site energy of carbon atoms of the top layer becomes different from that of the atoms of the bottom layer (see Fig.~\ref{f1}(a))~\cite{McCann2013Apr}.
This leads to the opening of a band gap at the two valleys, $K$ and $K'$ (see Fig.~\ref{f1}(b)), which depends on the strength of the symmetry breaking potential, i.e., on $D$ ~\cite{Oostinga2008Feb,Zhang2009Jun,Icking2022Nov}, resulting in a tunable band structure that allows for gate-defined charge carrier confinement~\cite{Eich2018Jul,Banszerus2020May, Banszerus2018Aug}.
The broken inversion symmetry also leads to a finite Berry curvature, $\Omega$, near the K points, where $\Omega$ has opposite signs at the $K$ and $K'$ points and has mirror symmetry for electrons and holes~\cite{McCann2013Apr, Banszerus2023Jun} (see Fig.~\ref{f1}(c)).
The Berry curvature gives rise to a valley-dependent anomalous velocity term leading to the valley-Hall effect in bulk BLG~\cite{Shimazaki2015Dec,Sui2015Dec} and to finite out-of-plane magnetic moments in BLG QDs. These topological orbital magnetic moments, which have opposite signs for $K$ and $K'$, couple to an external out-of-plane magnetic field and are the origin of the valley Zeeman effect in BLG QDs~\cite{Banszerus2021Sep,Tong2021Jan}.

To create gate-defined QDs in BLG, the electronic wave function needs to be confined by a potential $U(\textbf{r})$ in real space (see Fig.~\ref{f1}(a)) and will be distributed near the $K$ and $K'$ points, in k-space.
A single-electron or single-hole QD can then be described by the Hamiltonian $H_{\rm QD}=H_{\rm BLG}+H_{\rm Z}+H_{\rm SO}+U({\bf r})$~\cite{Wang2024Jul}. 
Here, $H_{\rm BLG}$ represents the effective $4 \times 4$ Hamiltonian of bulk BLG near the $K$- and $K'$-points based on the sublattice and layer degrees of freedom and includes the bulk valley Zeeman effect (see Appendix A), which will be further modified by the confinement $U({\bf r})$. $H_{\rm Z}$ denotes the spin Zeeman coupling.
$H_{\rm SO}$ describes the intrinsic Kane-Mele spin-orbit (SO) coupling, which lifts the zero $B$-field degeneracy of the four single-particle states, leading to the formation of two Kramers' pairs with opposite spin and valley quantum numbers, ($\ket{K \uparrow}$,$\ket{K' \downarrow}$) and ($\ket{K' \uparrow}$, $\ket{K \downarrow}$), separated by 
the SO gap, $\Delta_\mathrm{SO}$. In BLG devices $\Delta_\mathrm{SO}$ has typically values in the range of $40 - 80~\,\mu$eV~\cite{Banszerus2020May,Banszerus2021Sep,Banszerus2023Jun,Kurzmann2021Mar,Tong2022Feb}. 
In Fig.~\ref{f1}(d) we show the single-particle spectrum of a BLG hole QD as function of the out-of-plane magnetic field, $B_\perp$.
As $B_\perp$ couples to both the spin and the valley magnetic moments, we observe linear energy shifts given by $E(B_\perp)-E(0) = (\pm g_\mathrm{s} \pm g_\mathrm{v}) \mu_\mathrm{B} B_{\perp} / 2$~\cite{Knothe2018Oct}. 
Here, $\mu_\text{B}$ is the Bohr magneton, $g_\mathrm{s} \approx 2$ is the spin g-factor and the valley g-factor, $g_\mathrm{v}$, quantifies the strength of the Berry-curvature induced valley magnetic moment, which can be tuned by the confinement potential of the QD in a range typically between $g_\text{v} \approx 10$ and 70~\cite{Tong2021Jan,Moller2021Dec,Moller2023Sep}. 
All this makes the valley degree of freedom in BLG QDs highly tunable, in stark contrast to its behavior in Si, Ge, or SiGe QDs, enabling significant valley polarization at relatively low $B_\perp$-fields, as evident by the separation between the $\ket{K'}$ and excited $\ket{K}$ states shown in Fig.~\ref{f1}(d).

\begin{figure*}[ht]
\centering
\includegraphics[draft=false,keepaspectratio=true,clip,width=\linewidth]{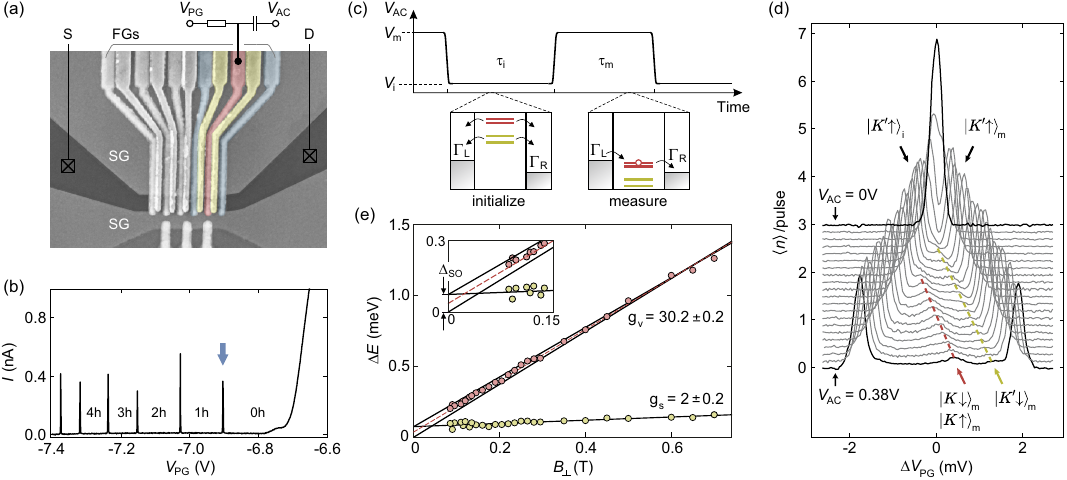}
\caption[Fig02]{
(a) False-colored scanning electron microscopy image of the gate structure of the device. The plunger gate (PG, red) is connected to a bias tee for applying AC and DC signals to the same gate.
(b) Current through the device as function of  $V_\text{PG}$ at a source-drain bias voltage of $V_\mathrm{SD}=200~\mu$V. The n-type channel is pinched off close to $V_\text{PG}=-6.8V$. Upon further decreasing $V_\text{PG}$, a hole QD is formed, and Coulomb resonances appear when additional holes are added to the QD (see labels).
(c) Top: Schematic of the square pulse applied to the PG characterized by the voltages $V_\text{i}$ and $V_\text{m}$ and the times $\tau_\text{i}$ and $\tau_\text{m}$. 
Bottom: Schematic of the QD states relative to the electrochemical potentials of the leads.
(d) Excited state (ES) spectroscopy using transient current measurements. The average number of charge carriers $\langle n \rangle$ tunneling through the QD per pulse is plotted as function of $\Delta V_\mathrm{PG}$ for different $V_\mathrm{AC}$ ($f = 5~$MHz, $B_\perp=300$~mT). Traces are offset for clarity.
$\ket{K'\uparrow}$ denotes current via the ground state. Orange and yellow dashed lines highlight transient currents via excited states ($\ket{K'\downarrow}$,  $\ket{K\uparrow}$ and $\ket{K\downarrow}$).
(e) Energy $\Delta E$ of the ES relative to the GS as a function of $B_\perp$. Fitting $\Delta_\text{SO}/2 + (g_\mathrm{v}+g_\mathrm{s}/2)\mu_\mathrm{B}B$ (orange dashed line) yields $g_\mathrm{v}$.
The solid lines indicate the energies of the valley ESs deduced from the fits.
The inset shows a close-up of the low-$B_\perp$ regime. }
\label{f2}
\end{figure*}

The fabricated device consists of a flake of BLG encapsulated by two crystals of hexagonal boron nitride (hBN) and placed on a graphite flake acting as a back gate. On top of the van-der-Waals heterostructure, split gates (SGs) are used to gap out the BLG underneath, resulting in a narrow n-type conductive channel connecting the source (S) and drain (D) leads (see Fig.~\ref{f2}(a)). 
To confine single charge carriers, the band edge profile along the channel can be adjusted using two layers of interdigitated finger gates (FGs)~\cite{Banszerus2018Aug,Eich2018Jul}. One of the FGs is used as a plunger gate (PG) to tune the QD, locally overcompensating for the channel potential set by the back gate. 
The width of the PG measures about 70~nm and the separation of the SGs is around 80~nm, setting an upper limit of the QD radius $r$ to around 30 to 40~nm. The DC potential applied to the plunger gate, $V_\text{PG}$, allows to control the charge carrier occupation down to the last hole (see Fig.~\ref{f2}(b)). 
To study transient transport through the QD, an AC potential, $V_\mathrm{AC}$, can be applied to the PG via a bias tee (see Fig.~\ref{f2}(a)). To maximize the transient currents and to study the relaxation dynamics of the QD states, the FGs adjacent to the PG (yellow and blue) are used to reduce the tunnel coupling between the QD and the left and right reservoir, $\Gamma_\mathrm{L}$ and $\Gamma_\mathrm{R}$~\cite{Hanson2003Nov,Banszerus2021Feb,Banszerus2022Jun,Gachter2022May}. Details on the device fabrication and the experimental setup are given in the Appendix B and C.

To study the relaxation dynamics of an excited valley state, we first investigate the single-particle spectrum of the QD. For that purpose, we perform excited state transient current spectroscopy measurements by applying a square pulse with a frequency, $f$, (duty cycle 50\%) to the PG (see Fig.~\ref{f2}(c)).
Fig.~\ref{f2}(d) shows the average number of charge carriers tunneling through the QD per pulse cycle, $\langle n \rangle /\text{pulse} = I/(fe)$, with the elementary charge $e$, as a function of the pulse amplitude, $V_\mathrm{AC}$, and the DC plunger gate voltage, $\Delta V_\mathrm{PG}$, relative to the Coulomb peak position at $V_\mathrm{AC} = 0$. 
At finite $V_\mathrm{AC}$, transport via the ground state (GS) may occur when the GS resides within the bias window during either part of the square pulse ($\tau_\mathrm{i}, \tau_\mathrm{m}$), resulting in a splitting of the GS Coulomb peak ($\ket{K \uparrow}_i$ and $\ket{K \uparrow}_m$)~\cite{Fujisawa2001Feb,Banszerus2021Feb}.
Once $V_\mathrm{AC}$ becomes large enough such that an excited state (ES) enters the bias window, a transient current via the ES contributes to the overall current and shows up as a resonance in Fig.~\ref{f2}(d) (see dashed lines). From the positions of the two prominent ES resonances we can extract their energies. 
Fig.~\ref{f2}(e) depicts the energy difference between the ground state ($\ket{K' \uparrow}$) and the first spin ES (yellow data points, $\ket{K'\downarrow}$) and the valley ESs, (red data points $\ket{K \downarrow}$, $\ket{K \uparrow}$), determined from measurements as in Fig.~\ref{f2}(d), as a function of $B_\perp$. 
The energy splitting of the spin ES and the GS increases linearly with $B_\perp$ due to the spin Zeeman effect. A fit to the data yields a spin g-factor of $g_\text{s}=2.0 \pm 0.2$ and a zero-field splitting of $\Delta_\text{SO}=75~\mu$eV, in agreement with the slightly proximity-enhanced Kane-Mele SO coupling~\cite{Banszerus2021Sep,Banszerus2023Jun,Kurzmann2021Mar}. 
Due to the finite peak width, the energy of the two valley ESs cannot be determined independently. Thus, the data has been fit considering the average energy splitting with a slope corresponding to $g_\mathrm{v}+g_\mathrm{s}/2$. 
A valley g-factor of $g_\mathrm{v}=30.2 \pm 0.2$ has been determined, similar to values reported in earlier works~\cite{Banszerus2021Sep,Tong2021Jan}.

\begin{figure}[ht]
\centering
\includegraphics[draft=false,keepaspectratio=true,clip,width=\linewidth]{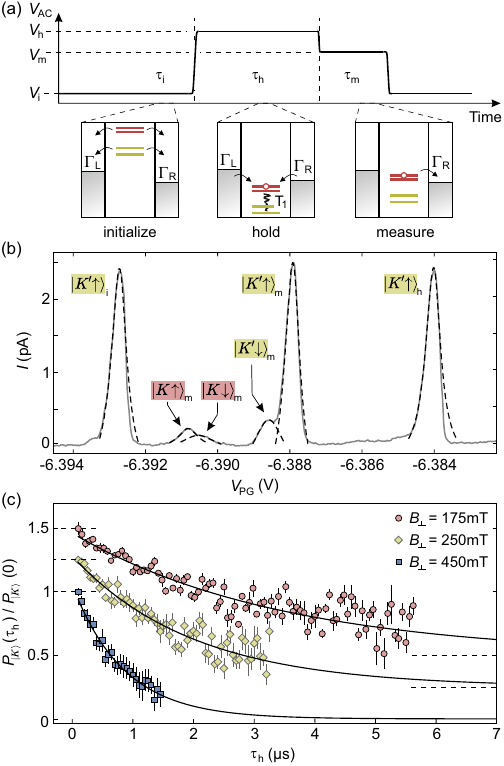}
\caption[Fig03]{
(a) Top: Schematic of the three-level pulse scheme applied to the PG which is characterized by the voltages $V_\text{i}, V_\text{h}$, $V_\text{m}$ and the times $\tau_\text{i}, \tau_\text{h}$ and $\tau_\text{m}$. 
Bottom: Schematic of the QD states relative to the electrochemical potentials in the leads (see text for details).
(b) Current $I$ as a function of $V_\text{PG}$ while the pulse sequence in \textbf{a} is applied ($B_\perp=0.22$~T, $V_\mathrm{SD}=10~\mu$V). The valley $T_1$ time is derived from the amplitude of $\ket{K \uparrow}_\text{m}$. Dashed curves are Lorentzian fits to the peaks.
(c) Relative occupation probability of $\ket{K\uparrow}$ after the holding pulse, $P_{\ket{K}} (\tau_\text{h})/P_{\ket{K}} (0)$ as a function of the holding time, $\tau_\text{h}$. The traces are offset for clarity.}
\label{f3}
\end{figure}

Next, to investigate the relaxation dynamics of the observed single-particle valley ES, we apply a three-level pulse scheme to the PG (see Fig.~\ref{f3}(a)) and measure the tunneling current through the QD~\cite{Banszerus2022Jun}. 
During $\tau_\mathrm{i}$, the QD is initialized in the empty state. 
Subsequently, during $\tau_\mathrm{h}$, the GS $\ket{K'\uparrow}$, the spin ES $\ket{K'\downarrow}$ and the two valley ESs $\ket{K\uparrow}$ and $\ket{K\downarrow}$ are tuned below the electrochemical potentials of the source and drain leads. After the characteristic tunneling time, on the order of $1/(\Gamma_\mathrm{L}+\Gamma_\mathrm{R})$ either the GS or one of the three ESs will be occupied by a single charge carrier.
A charge carrier in an ES has the chance to relax into an energetically lower lying state by either spin or valley relaxation with a characteristic relaxation time, $T_1$.  
Finally, during $\tau_\mathrm{m}$, we perform a valley selective readout, measured by aligning the $\ket{K}$ states in the bias window. 
Only charge carriers occupying one of the two $\ket{K}$ states, which have not relaxed into a $\ket{K'}$, can tunnel out and contribute to the transient current.
Fig.~\ref{f3}(b) shows the current, $I$, through the QD as a function of $V_\mathrm{PG}$ while applying the pulse sequence depicted in Fig.~\ref{f3}(a). 
The three peaks labeled $\ket{K'\uparrow}_\mathrm{i}$, $\ket{K'\uparrow}_\mathrm{h}$ and $\ket{K'\uparrow}_\mathrm{m}$ correspond to GS transport during each of the three pulse steps.
Furthermore, transient currents via the three ESs, $\ket{K\uparrow}_\mathrm{m}$, $\ket{K\downarrow}_\mathrm{m}$ and $\ket{K'\downarrow}_\mathrm{m}$, can be observed during $\tau_\mathrm{m}$.
The relaxation time $T_1$ of the $\ket{K}$ states into an energetically lower lying state can be probed by extracting the maximum value of the combined $\ket{K \uparrow}_\mathrm{m}$ and $\ket{K \downarrow}_\mathrm{m}$ peaks, which at a fixed value of $B_\perp$ is at a constant energy difference (see Fig.~\ref{f1}(d)), as a function of the holding time $\tau_\mathrm{h}$ (for more information on the peak analysis see Appendix D)~\cite{Hanson2003Nov,Banszerus2022Jun,Gachter2022May}. 
We convert the current $I$ into the number of charge carriers tunneling through the QD per pulse cycle, $\langle n \rangle /\text{pulse} = I(\tau_\text{i}+\tau_\text{h}+\tau_\text{m})/e$.
The number of charge carriers $\langle n \rangle_{\ket{K}_\mathrm{m}}$ tunneling via the excited $\ket{K}$ states, is directly proportional to the probability of $\ket{K}$ remaining occupied after $\tau_\text{h}$, $P_{\ket{K}} (\tau_\text{h})$. 
The relative occupation probability of $\ket{K}_\mathrm{m}$ as a function of $\tau_\mathrm{h}$ decays exponentially with the characteristic relaxation time, $T_1$, 
$\langle n \rangle_{\ket{K}_\mathrm{m}}(\tau_\mathrm{h})/\langle n \rangle_{\ket{K}_\mathrm{m}}(0)=P_{\ket{K}}(\tau_\text{h})/P_{\ket{K}}(0)=e^{-\tau_\mathrm{h}/T_1}$~\cite{Hanson2003Nov,Banszerus2022Jun}.

\begin{figure}[t]
\centering
\includegraphics[draft=false,keepaspectratio=true,clip,width=0.98\linewidth]{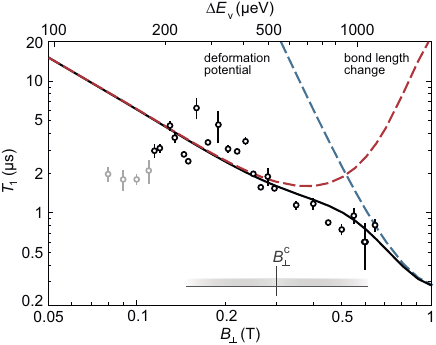}
\caption[Fig04]{
Valley relaxation time $T_1$ as a function of $B_\perp$ (bottom axis) and the valley splitting $\Delta E=~g_\text{v}\mu_\text{B} B_\perp$ (top axis). Error bars indicate the 1$\sigma$ confidence interval of an exponential fit to the data as in Fig.~\ref{f3}(c).
The black curve represents a fit assuming $T_1$ to be limited by electron-phonon coupling arising from the deformation potential and from bond length change. The blue (red) curve shows the contribution of the deformation potential (bond length change) separately.
}
\label{f4}
\end{figure}

Fig.~\ref{f3}(c) depicts $P_{\ket{K}}(\tau_\text{h})/P_{\ket{K}}(0)$ as function of $\tau_\mathrm{h}$ for three different out-of-plane magnetic fields.
The datasets show an exponential decay of the occupation probability as a function of $\tau_\mathrm{h}$. 
An exponential fit (solid line), yields for example $T_1 = 4.0~\mu$s at $B_\perp=0.175$~T. 
$T_1$ decreases with increasing $B_\perp$ and reaches a value of 845~ns at $B_\perp=0.45$~T. 
A single charge carrier occupying $\ket{K \uparrow}$ or $\ket{K\downarrow}$ may relax into a lower lying state either by pure valley relaxation ($\ket{K \uparrow} \rightarrow \ket{K' \uparrow}$ and $\ket{K\downarrow} \rightarrow \ket{K'\downarrow}$) or by additionally flipping the spin ($\ket{K \uparrow} \rightarrow \ket{K' \downarrow}$ and $\ket{K\downarrow} \rightarrow \ket{K'\uparrow}$). 
Relaxation processes requiring a single valley flip are expected to be faster than processes that require both a spin and valley flip. 
This is supported by recently reported spin relaxation times between hundreds of microseconds to up to 50~ms, for energy splittings, $\Delta E_\mathrm{S} > 200$~$\mu$eV~\cite{Banszerus2022Jun,Gachter2022May}. 
For the pure spin relaxation between $\ket{K'\downarrow}$ and the GS, no relaxation could be observed over the whole range of investigated $\tau_\mathrm{h}$ and $B_\perp$ (the amplitude of $\ket{K'\downarrow}_\mathrm{m}$ is constant as a function of $\tau_\mathrm{h}$, see Appendix D).
Hence, we conclude that $T_1$ extracted from Fig.~\ref{f3}(c) must be limited by the valley relaxation time.\\
In Fig.~\ref{f4}, we plot the valley relaxation time $T_1$ extracted from exponential fits, as exemplarily shown in Fig.~\ref{f3}(c) as a function of $B_\perp$ and as a function of the energy splitting $\Delta E_\mathrm{v}$.
When decreasing $B_\perp$ from 0.7~T to about 0.15~T, $T_1$ increases from below 0.5~$\mu$s to about 7~$\mu$s, while at even lower $B_\perp$, the relaxation rate decreases again to $T_1 \sim$ 2~$\mu$s at 80~mT (gray).

To gain a better understanding of the experimental results, we compare them with theory. We model the system by the Hamiltonian $H=H_{\rm QD}+H_{\rm EPC}+H_{KK'}$, where
$H_{\rm QD}$ describes a single electron or hole in the BLG QD and $H_{\rm EPC}=\sum_{\lambda q}H_{\rm EPC}^{\lambda q}$ the electron-phonon coupling.
Furthermore, we allow a mixing between the two valleys described by the intervalley coupling term, $H_{KK'}=\Delta_{KK^{\prime}}\tau_x/2$ with the Pauli matrix $\tau_x$ acting on the valley degree of freedom. 
For simplicity, we model electrostatic confinement by a finite circularly symmetric step potential $U({\bf r})$ with potential depth $U_0 \approx 39.6\ $meV and  $r=25~$nm.
This yields a valley g-factor of $g_\text{v}=30$, in good agreement with the experiment (c.f. Fig.~\ref{f2}(e)).
We consider transitions between states with equal spin, but opposite valley degree of freedom mediated by coupling to in-plane acoustic phonons arising either from the deformation potential (coupling strength $g_1$) or from bond-length change ($g_2$)~\cite{Sohier2014Sep}.
Since we operate in the low energy limit, we only consider acoustic phonons, while out-of-plane acoustic (ZA) phonons are supposed to be quenched in supported, especially in encapsulated graphene~\cite{Struck2010Sep,Wang2024Jul}.
The Hamiltonian describing coupling to phonons in the mode $\lambda$ with wave vector $q$ has the form 
$H_{\rm EPC}^{\lambda q}=c_q(g_1a_1\sigma_0+g_2a'_2\sigma_x+g_2a''_2\sigma_y)(e^{i{\bf q}\cdot{\bf r}}b^{\dagger}_{\lambda q}-e^{-i{\bf q}\cdot{\bf r}}b_{\lambda q})$ with $\sigma_{x,y,z}$ the Pauli matrices for the sublattice 
degree of freedom~\cite{Ando74_2005,Mariani100_2008}, and $c_q=\sqrt{q/A\rho v_{\rm \lambda}}$, with $A$ the area of the BLG sheet, $\rho$ the mass density of BLG, and $v_{\rm \lambda}$ the sound velocity; $a_{1,2}$ are phase factors and $b_{\lambda q}$ and $b^\dagger_{\lambda q}$ are the phonon ladder operators~\cite{Wang2024Jul}.   
Using Fermi's golden rule, we calculate the valley relaxation times $T_1$ between initial and final eigenstates $\ket{i}$ and $\ket{f}$ of the Hamiltonian $H_{\rm QD} + H_{KK'}$ with opposite valley quantum number and eigenenergies $\varepsilon_i$ and $\varepsilon_f$,
\begin{equation*} \label{GoldenRule}
    \frac{1}{T_1} = 2\pi A \sum_\lambda \int \frac{d^2q}{(2\pi)^2} |\bra{i}H_\mathrm{EPC}^{\lambda q}\ket{f}|^2 \delta(\varepsilon_f - \varepsilon_i + v_\mathrm{\lambda} q) .
\end{equation*}
We only take into account the emission of phonons (with energy $v_\mathrm{\lambda} q$) as the thermal energy is significantly smaller than the valley splitting.
To quantify the electron-phonon coupling strength, we perform the least square fit to the experimental data using $g_1$ and $g_2$ as free fit parameters. 
Our model is in good qualitative and quantitative agreement with the data taken above $B_\perp = 0.1$~T, where increasing $B_\perp$ results in decreasing $T_1$, while it cannot explain the decrease in $T_1$ observed for $B_\perp<0.1~$T, suggesting that other mechanisms dominate in this regime.
We speculate that the discrepancy between our model and the data in the regime $B_\perp<0.1~$T may be due to a hot spot~\cite{Yang2013Jun}, thermal broadening induced charge noise~\cite{Huang2014May, Hosseinkhani2021Aug} or $1/f$ charge noise, which is expected to become dominant in the small energy range, as discussed in more detail in ref.~\cite{Wang2024Jul}.
Consequently, we have restricted the fit to the data for $B_\perp \geq 0.1$~T, which yields coupling parameters of $g_1=50~$eV, $g_2=5.4~$eV. 
It is noteworthy that both parameters are in good agreement with literature, which includes values in the range of $20$ to $50$~eV for $g_1$~\cite{Park2014Mar,Mariani2010Nov,Chen2008Apr,Hwang2008Mar,Suzuura2002May} and values in the range of $1.5$ to $5$~eV for $g_2$~\cite{Suzuura2002May,Sohier2014Sep},
where the wide range of values is partly due to the dependence of the deformation potential on screening and doping~\cite{Park2014Mar,Mariani2010Nov}.
The black solid line in Fig.~\ref{f4} corresponds to the contributions from both, the deformation potential coupling and from bond length change coupling, while the dashed lines show the individual contributions, respectively (see labels).
In the calculation, the intervalley coupling, which is mainly responsible for the absolute values of $T_1$ but does not enter the functional $B_{\perp}$ dependence, was set to $\Delta_{KK'} = 50~\mu$eV. 
At larger magnetic fields, $B_\perp \gtrsim 0.5~$T, $T_1$ is predominantly limited by electron-phonon coupling via the deformation potential, whereas at smaller fields it is limited by the coupling due to bond length change. 
This transition occurs due to the crossover between the dipole and the higher multipole regimes for the bond-length change coupling if the phonon wavelength $\lambda\approx 2\pi\hbar v_{\rm \lambda}/(g_{\rm v}\mu_B B_{\perp})$ is comparable to the QD radius $r$, where $q r = 2\pi r/\lambda\approx 1$.
Hence, the crossover occurs around the critical field $B^{c}_{\perp}\approx \hbar v_{\rm \lambda}/(g_{\rm v}\mu_\mathrm{B} r)  \approx$ 0.3~T.
The assumed QD radius of $r = 25$~nm is in agreement with the lithographic device dimensions as well as with the confinement size, giving rise to a valley g-factor of $g_\text{v}=30$ in excellent agreement with experiment. 
The gray bar in Fig.~\ref{f4} depicts the range of $B^{c}_{\perp}$ assuming the estimate of $r$ to deviate by a factor of 2 highlighting that the transition region is well within the experimentally investigated $B_\perp$ range. 
The long single-particle valley lifetimes in BLG QDs of up to 7~$\mu$s make BLG a promising candidate for valleytronic applications and confirm that the valley degree of freedom is indeed interesting for the realization of qubits, where the $T_1$ time sets an upper limit on the coherence time $T_2^*$.

This potential is furthermore underlined by a recent experiment showing long relaxation times from valley triplet to valley singlet states~\cite{Garreis2024Jan}.
Fits to the experimental data confirm that electron-phonon coupling mediated by bond length change and the deformation potential limit the relaxation time over a wide magnetic field range. 
As the valley magnetic moment is typically one to two orders of magnitude larger than the magnetic moment associated with the electron spin, we anticipate gate operation times of a valley qubit to be much faster than those of a spin qubit, potentially compensating for the shorter relaxation times. 
The magnitude of the valley magnetic moment can be adjusted all-electrically~\cite{Moller2023Sep}, which could provide a way to realize control over a single valley without the need for microwave bursts, micromagnets or ESR strips, and enable well-controlled addressability. 
Crucial follow-up experiments include the determination of the coherence times ($T_2^*$ and $T_2$) and insights into their limiting mechanisms such as charge noise, as well as the understanding of the upper bound of the recently reported long relaxation times of Kramers states at small energy splittings~\cite{Denisov2025Apr}.\\

\textbf{Acknowledgements}\\
The authors thank A. Hosseinkhani for fruitful discussions and F. Lentz, S. Trellenkamp, M.~Otto and D. Neumaier for help with sample fabrication. This project has received funding from the European Research Council (ERC) under grant agreement No. 820254, the Deutsche Forschungsgemeinschaft (DFG, German Research Foundation) under Germany's Excellence Strategy - Cluster of Excellence Matter and Light for Quantum Computing (ML4Q) EXC 2004/1 - 390534769, through DFG (STA 1146/11-1), and by the Helmholtz Nano Facility~\cite{Albrecht2017May}. K.W. and T.T. acknowledge support from the JSPS KAKENHI (Grant Numbers 21H05233 and 23H02052) , the CREST (JPMJCR24A5), JST and World Premier International Research Center Initiative (WPI), MEXT, Japan.
L.W. and G.B. acknowledge support from DFG Project No.~425217212, SFB 1432.\\\\
\textbf{Data availability}\\
The data supporting the findings are available in a Zenodo repository under accession code XXX. \\\\
\textbf{Author contributions}\\
C.S. designed and directed the project; L.B., K.H., S.M. fabricated the device, L.B., K.H. and C.V. performed the measurements and analyzed the data. K.W. and  T.T.  synthesized  the  hBN  crystals. L.W. and G.B. performed calculations of the T$_1$ time.  G.B., C.V. and C.S. supervised the project. L.B., K.H., L.W., G.B, C.V. and C.S. wrote the manuscript with contributions from all authors. L.B. and K.H. contributed equally to this work. Correspondence should be addressed via e-mail to katrin.hecker@rwth-aachen.de \\\\
\textbf{Competing interests}\\
The authors declare no competing interests.

\section{Appendix}
\subsection{A. Hamiltonian of bilayer graphene}
The Hamiltonian $H_\mathrm{BLG}$ used to describe the band structure of bulk BLG is given by
\begin{equation}
    H_\mathrm{BLG}(\mathbf k) = 
    \begin{bmatrix}
        \Delta & \gamma_0p & \gamma_4p* & \gamma_1\\
        \gamma_0p* & \Delta & \gamma_3p & \gamma_4p*\\
        \gamma_4p & \gamma_3p* & -\Delta & \gamma_0p\\
        \gamma_1 & \gamma_4p & \gamma_0p* & -\Delta
    \end{bmatrix}
\end{equation}
with the displacement field $2\Delta$ and the hopping parameters $\gamma_0 = 2.6$~eV, $\gamma_1 = 0.339$~eV, $\gamma_3 = 0.28$~eV and $\gamma_4 = -0.14$~eV. The momentum $p(\mathbf{k}) = -\sqrt{3}a (\tau k_x -ik_y -ixB_\perp e/2 -\tau yB_\perp e/2)/2$ with the valley index $\tau = \pm1$ and the lattice constant $a = 2.46$~Å includes the valley Zeeman effect~\cite{McCann2013Apr, Konschuh2012Mar, Wang2013May}.\\

\subsection{B. Device fabrication}
The device is composed of a van-der-Waals heterostructure, where a BLG flake is encapsulated between two hBN flakes of approximately 25~nm thickness and placed on a graphite flake which acts as a back gate (BG). Cr/Au split gates on top of the heterostructure define a 80~nm wide channel. Across the channel, two layers of interdigitated Cr/Au finger gates of 70~nm width, are fabricated. Two 15~nm thick layers of atomic layer deposited (ALD) Al$_2$O$_3$ act as gate dielectric. For details of the fabrication process, we refer to Ref.~\cite{Banszerus2020Oct}.

\subsection{C. Experimental setup and parameters}
In order to perform RF gate modulation, the sample is mounted on a home-built printed circuit board (PCB). All DC lines are low-pass-filtered (10~nF capacitors to ground). All FGs are connected to on-board bias-tees, allowing for AC and DC control on the same gate (see Fig.~2a of the manuscript). All AC lines are equipped with cryogenic attenuators of -26~dB. $V_\text{AC}$ refers to the AC voltage applied prior to attenuation. The measurements are performed in a $^3$He/$^4$He dilution refrigerator at a base temperature of approximately 20~mK and at an electron temperature of around 60~mK. The current through the device is amplified and converted into a voltage with a home-built I-V converter at a gain of $10^8$.

Throughout the experiment, a constant back gate voltage of $V_\mathrm{BG} = 5.025~$V and a split gate voltage of $V_\mathrm{SG} = -5.435~$V is applied to define a n-type channel between source and drain.
The four gates acting as barrier gates to the QD (see Fig.~2a of the manuscript, yellow and blue color coding) are biased with $-6.05 \pm 0.1$~V, -4.95~V, -5.18~V and $-6.15 \pm 0.1$~V, respectively. The voltages are adjusted to compensate for the influence of $B_{\perp}$ on the tunnel coupling.

\subsection{D. Peak analysis and extracting relaxation times}
\begin{figure*}[]
\centering
\includegraphics[draft=false,keepaspectratio=true,clip,width=0.6\linewidth]{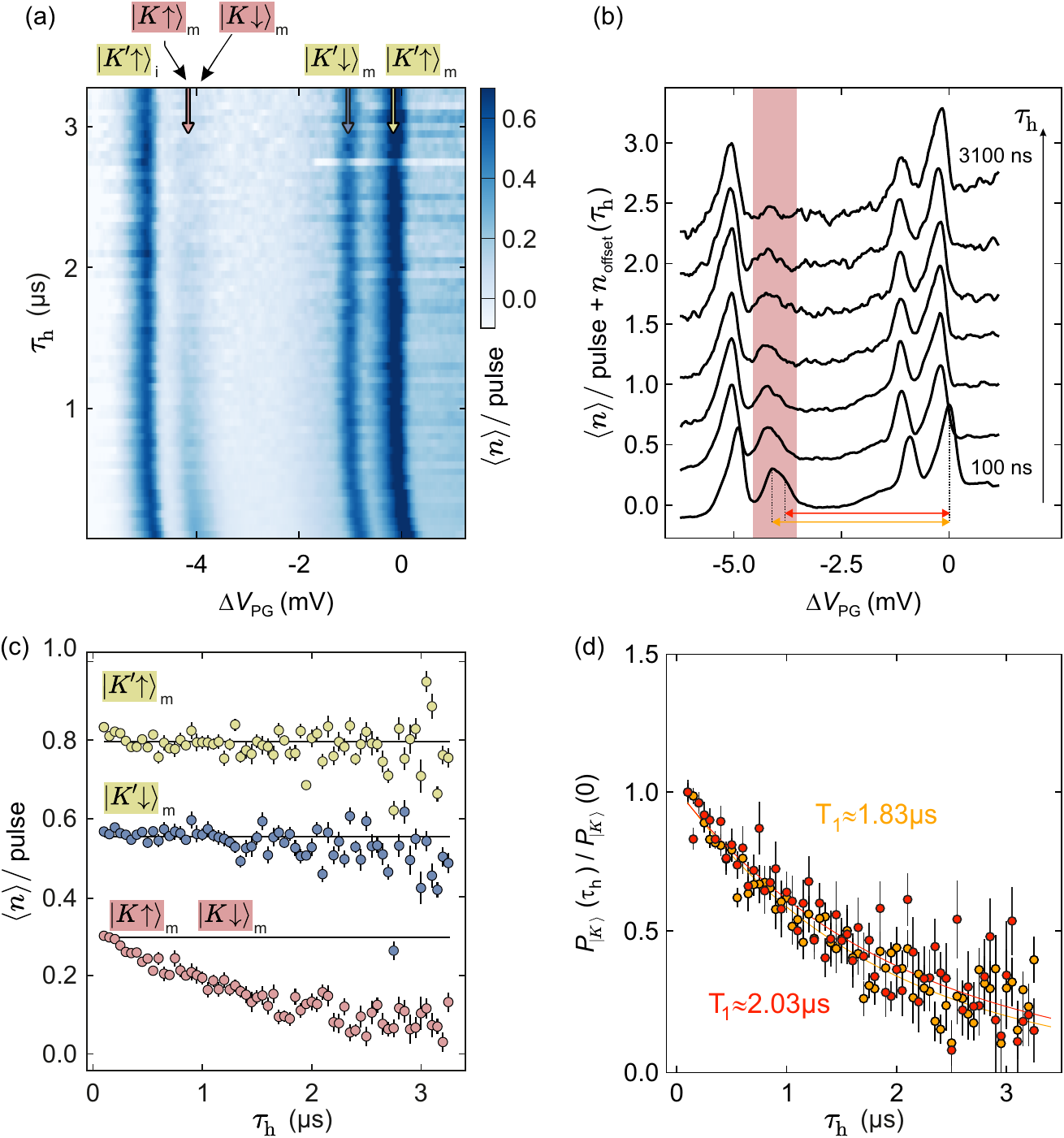}
\caption[FigS1]{(a) Average number of charge carriers, $\langle n \rangle$, tunneling through the device per pulse cycle (derived from the current $I$) as a function of $V_\mathrm{PG}$ and $\tau_\mathrm{h}$ while the pulse sequence depicted in Fig.~\ref{f3}(a) is applied. The color coded arrows mark the positions of the peaks $\ket{K'\uparrow}_\mathrm{m}$ (yellow), $\ket{K'\downarrow}_\mathrm{m}$ (blue) and $\ket{K\downarrow}_\mathrm{m}$, $\ket{K\uparrow}_\mathrm{m}$ (red). An out-of-plane magnetic field of $B_\perp=0.25$~T was applied. 
(b) Exemplary line traces of the measurement in (a), where $\langle n \rangle /\mathrm{pulse}$ is shown as function of $V_\mathrm{PG}$ for different values of $\tau_\mathrm{h}$ (black arrow on the right side). The traces are offset by an arbitrary $n_\mathrm{offset}$ for better visibility. 
(c) Maximum values of the labeled peaks in (a) as a function of the pulse duration $\tau_\mathrm{h}$. The maximum value of the peak that we attribute to $\ket{K\downarrow}_\mathrm{m}$ and $\ket{K\uparrow}_\mathrm{m}$ was extracted in a voltage window at a fixed energy difference from the peak $\ket{K'\uparrow}_\mathrm{m}$ (red shaded area in (b)). 
(d) Normalized occupation probability extracted at fixed energy splittings corresponding to the two $\ket{K}$ states, respectively (red and orange arrows in (b)).}
\label{fS1}
\end{figure*}

Fig.~\ref{fS1}(a) shows an exemplary measurement of the average number of charge carriers, $\langle n \rangle$, tunneling through the QD per pulse cycle as a function of the duration of the holding pulse, $\tau_\mathrm{h}$, and the plunger gate voltage $V_\mathrm{PG}$. The colored labels mark the probed states. 
The measurement was recorded at an out-of-plane magnetic field of $B_\perp=0.25$~T (measurement corresponds to the results depicted in Fig.~\ref{f3}(c)). Exemplarily, line cuts of the measurement in Fig.~\ref{fS1}(a) are shown in Fig.~\ref{fS1}(b). 
The peak heights of $\ket{K'\uparrow}_\mathrm{m}$ and $\ket{K'\downarrow}_\mathrm{m}$ are displayed in Fig.~\ref{fS1}(c) as function of the pulse duration $\tau_\mathrm{h}$. The maximum value of the peak attributed to $\ket{K\uparrow }_\mathrm{m}$ and $\ket{K\downarrow }_\mathrm{m}$ (red data points in Fig.~\ref{fS1}(c)) is extracted in a small voltage window at a fixed distance from the position of the peak $\ket{K'\uparrow }_\mathrm{m}$ (visualized by the red shaded area in Fig.~\ref{fS1}(b)). 

It is important to note that the observed exponential decay does not depend on details of the peak analysis. The line cuts in Fig.~\ref{fS1}(b) reveal, that the finite peak linewidth compared to the small energy spacing between the $\ket{K}$ states makes it difficult to identify the peaks $\ket{K\downarrow }_\mathrm{m}$ and $\ket{K\uparrow }_\mathrm{m}$ independently. Therefore, we extracted the maximum value of the combined peak contributions (red shaded area in Fig.~\ref{fS1}(b)). 
To highlight that the obtained result is robust, we determine the values at fixed energy splittings from the ground state peak $\ket{K'\uparrow }_\mathrm{m}$, $\Delta E_\mathrm{V}+\Delta E_\mathrm{S}$ (red arrow in Fig.~\ref{fS1}(b)) and $\Delta E_\mathrm{V}+\Delta_\mathrm{SO}$ (orange arrow in Fig.~\ref{fS1}(b)), which correspond to the expected positions of the peaks $\ket{K\downarrow }_\mathrm{m}$ and $\ket{K\uparrow }_\mathrm{m}$, respectively. The corresponding normalized occupation probabilities as function of $\tau_\mathrm{h}$ are shown in Fig.~\ref{fS1}(d). The results yield similar decay constants ($T_1\approx2.03\,\mu$s and $T_1\approx 1.83\,\mu$s) which are within the margin of the error of the results shown in Fig.~\ref{f4}.
Apart from the exponential decay of the maximum value at $\ket{K\downarrow }_\mathrm{m}$ and $\ket{K\uparrow }_\mathrm{m}$ (see also Fig.~\ref{f3}c), the peaks of the ground state ($\ket{K'\uparrow}_\mathrm{m}$) and the first excited state ($\ket{K'\downarrow}_\mathrm{m}$) show a constant maximum value over the measured range of $\tau_\mathrm{h}$. 
Since spin relaxation, from $\ket{K'\downarrow}$ to $\ket{K'\uparrow}$, would show up in a decrease of the peak height of $\ket{K'\downarrow}_\mathrm{m}$ (blue data points in Fig.~\ref{fS1}(c)), we conclude that spin relaxation appears on a timescale $\tau_\mathrm{h}\gg3\,\mathrm{\mu s}$. This observation aligns with previous findings \cite{Banszerus2022Jun,Gachter2022May}.


%

\end{document}